# Domain Specific Question to SQL Conversion with Embedded Data Balancing Technique

Jyothi[1], T.Satyanarayana Murthy[2]

[1]jotisuv@gmail.com, [2]tsmurthy_it@cbit.ac.in
[1]Data Science, LJMU/UPGRAD, Street-1, Bengaluru, Karnataka
[2]Department of IT, Chaitanya Bharathi Institute of Technology(Autonomous), Hyderabad, Telangana, India

**ABSTRACT**

The rise of deep learning in natural language processing (NLP) has fostered the creation of text to structured query language (SQL) models composed of an encoder and a decoder. Researchers have experimented with various intermediate processing like schema linking, table type aware, value extract etc. to generate accurate SQL results for the user question. However, error analysis performed on the failed cases on these systems shows, 29% of the errors would be because the system was unable to understand the values expressed by the user in their question. This challenge affects the generation of accurate SQL queries, especially when dealing with domain specific terms and specific value conditions, where traditional methods struggle to maintain consistency and precision. To overcome these obstacles, this thesis proposes two intermediations: implementing data balancing technique like over sampling domain specific queries which would refine the model's architecture to enhance value recognition and fine tuning the model for domain specific questions. This proposed solution demonstrated 10.8% improvement in accuracy of the model performance compared to the state-of-the-art model tested on WikiSQL dataset. to convert the user question accurately to SQL queries. Applying oversampling technique on the domain specific questions shown a significant improvement in the model's ability

INTRODUCTION

1.1    Background

With the growing digitization, relational databases are widely used to store the business data. This data can be accessed by the user, using SQL. User wants to see this data or needs additional information on the available subject. To do that, the user would need to know the skill of querying these databases using SQL. This causes limitations to the users, that the data cannot be reviewed by them easily when needed. To make data accessible to everyone, research has been performed in developing text to SQL conversion models. With the popularity of deep learning models, there are several systems built using deep learning-based approaches for converting user questions into SQL query. The text to SQL model enables the users to easily search databases using Natural Language Question (NLQ). Also, SQL operators like MIN, MAX, COUNT, and keywords like JOIN, GROUP BY are used to extract information from the database. Thus, text to SQL helps non-technical users in getting required information. Text to SQL conversion is applicable in diverse fields like virtual assistants, chatbot etc. Text to SQL is one of the important areas of study in NLP, and different approaches have been explored for different purposes. With the rise of deep learning in NLP, there are a lot of systems developed using deep learning models (Yu et al., n.d.; Zhong et al., 2017; Lyu et al., 2020; Mellah et al., 2020; Pal et al., 2021; Xu et al., 2021). The model uses encoder and decoder, where the encoder converts input into vectors and the decoder uses these vectors to convert user questions into an SQL query. Decoders can be of three kinds [7] sketch-based, which views SQL query components as slots to be filled, and sequence-based, which sequentially generates the query, and grammar based, which generates a grammatical rule. Schema linking is one of the areas explored by many systems to extract the database related information from the natural language (NL) questions. The "Schema Linking" section referenced in the work [7] discusses the process of connecting elements of a NL question to relevant components of a database schema. The work mentions candidate discovery and candidate matching. Candidate discovery is the identification of query candidates from the NL question and database candidates from the database. Candidate matching then compares these sets to establish links. [8], introduces IRNet, which tackles the disparity between spoken language intent and SQL query specifics. It emphasizes schema linking, SemQL query synthesis, and SQL query inference. [9] also focuses on incorporating database content and metadata into SQL queries. It presents two systems, ValueNet Light and ValueNet, employing advanced pre-processing and neural network architectures. The work addresses challenges in column prediction and value incorporation in SQL queries. However, suggests future improvements in architecture and generative neural network approaches for good value candidate prediction. The modified IRNet [10] enhances IRNet by incorporating data oversampling, revising SemQL grammar and modifying the loss function to improve the accuracy of text-to-SQL translation, especially for nested SQL queries. This work suggests enhancing the loss function further to improve overall accuracy on nested queries. [11] employs relation aware self-attention technique to enhance the text to SQL conversion systems ability to understand and encode complex relationship in the database. The model uses unified framework that includes schema encoding, schema linking, and feature representation within a text-to-SQL encoder. However, when reviewing the errors the system struggles to handle the domain specific queries. One of the observations from above research work is missing where clause specially when the questions include domain specific phrases like smaller than, less than or bigger than etc. This thesis would be referencing the questions with term smaller than, bigger than, larger than conditions as domain specific query for the ease of classification. As per IRNet error analysis, 14% of the errors were due to these type of user questions. Similarly, RAT-SQL reported 29% of the errors were due to missing where clause. The reason for these questions resulting in missing where clause could be due to limited dataset of these kind of queries. Hence, this can be considered as imbalanced data and can be attempted to solve by experimenting data balancing techniques. This research work is trying to address issues by recommending two changes. Implement the data balancing technique to over sample the domain specific queries in the WikiSQL dataset to enhance value recognition and fine-tune the model on subset data. This combination helps in minimising errors by focusing on correctly predicting individual components of the query. By focusing on these experiments, this research aims to significantly bridge the gap between natural language questions and their corresponding SQL representations, leading to robust and user-friendly database interaction systems. In summary, this would result in enhanced adaptability of the text to SQL conversion systems.

1.2    Aims and Objectives

The aim of this research work is to explore ways to increase the accuracy of Text to SQL Conversion. The goal of the research is to develop a better process to improve value candidate generation through data balancing technique on domain specific questions. This will result in increase in the accuracy of the model.

Objectives
- To develop the proposed architecture to come up with good value candidates from the natural language by introducing data balancing technique

- To fine tune the model to improve its ability to understand and process domain-specific words accurately.
- To evaluate the model on a domain-specific dataset to ensure the model accuracy improved.

LITERATURE REVIEW

2.1     Introduction

Literature review for this work was started initially using survey literature available in this field. [12] discusses the increasing complexity, volume of data collected, and the need for systems with high interactivity and effective solutions for data retrieval. This research work delves into advancements in data warehousing to enhance extract, transform, and load (ELT) efficiency and facilitate efficient data retrieval. The work also explores recent innovations in data warehousing technology, including multi-disk buffering during the ELT process and the emergence of hybrid multi-dimensional and relational online analytical processing (OLAP) databases (HOLAPs). The objective is to enhance data access, retrieval, and reporting efficiency in a customer-facing context. [13] details the creation of a text to SQL dataset and explores various technologies related to the task model and its practical applications. It identifies areas for enhancement in large-scale annotated datasets like WikiSQL [1] and Spider [14], while also highlighting the scarcity of research on Chinese language. The work outlines the evolution of Text to SQL datasets from single-domain to cross-domain and from single-table mode to multi-table mode. Various decoding methods, including intermediate representation, tree decoding, and graph network modelling, are summarized, with varying effectiveness for complex data. [15]contributes to the on-going dialogue in the field by conducting a comprehensive comparative analysis of various deep learning methods used for text to SQL conversion. The comparison of different models, including sequence-to-sequence and slot-filling methods, provides valuable insights into the strengths and weaknesses of current methodologies. This work emphasizes the difficulties in creating text to SQL models that are both scalable and accurate for a wide range of database schemas, including those that are unfamiliar. [7]provided an extensive overview of the present landscape of deep learning text to SQL systems, covering benchmarks, evaluation methods, and recent advancements. The work introduces a taxonomy that enables direct comparisons between different approaches and categorizes techniques for each translation step. Additionally, it evaluates the pros and cons of design choices within these systems, leveraging the taxonomy as a guiding framework. Lastly, the work outlines open challenges and suggests critical directions for future research in the field of text to SQL systems. Using above information, the literature review will first provide details on the dynamic field of converting natural language questions into SQL queries. The section is broken down into various sections. It will outline the two main types of Text-to-SQL datasets: single domain datasets like IMDb, multi-domain datasets like WikiSQL [1] and Spider [14]. The workings of these datasets, including their domain specifics and query complexity, are reviewed. Then it will provide a review on NL questions, highlighting the importance of understanding the user's intent and the specific components of a question, such as identifying relevant columns and tables. Key studies in this domain are analysed, including advancements in deep learning text-to-SQL systems and the challenges of lexical, syntactic, and semantic ambiguities in natural language questions. These studies provide insights into the evolution of Text-to-SQL datasets, decoding methods, and the role of pre-trained language models (PLMs) in enhancing natural language question interpretation. The review scrutinizes various systems developed for natural language to SQL translation, examining their methods, challenges, and potential future enhancements. The focus is on improving the accuracy and efficiency of these systems.

## 2.2 Dataset

Text to SQL dataset encompasses a collection of natural language questions paired with their corresponding SQL queries, defined across one or more databases. The datasets are broadly of two types: one domain dataset and multiple domain dataset. Dataset which are specific to one domain have examples for one specific area like IMDB which is specific to movies or television shows. Other examples of datasets specific to one domain are GeoQuery, ATIS, and Restaurants. They perform well on their own domain however they cannot perform when faced with unseen question. Datasets with more than one domain are large and they are widely used in the text to SQL conversion research. WikiSQL (Zhong et al., 2017a) and Spider (Yu et al., 2018c) are the two large scale multi-domain datasets which can be used to train and test the different neural network techniques and compare them against different systems. WikiSQL introduced with seq2SQL [1] system and is one of the largest datasets for text to SQL conversions. This includes 25,000 Wikipedia tables and 80000+ NL and SQL question pairs. The query complexity in this dataset is less as they are all simple SQL statements with single database. They do not include SQL clauses like joins, group by or order by. Spider was created by 11 college students, and it addresses the absence of complex, cross domain dataset. It includes 200 databases with more than one table, 10000+ questions and 5000+ SQL queries for these questions. There are other multi domain datasets like SQUALL which includes natural language questions on Wikipedia tables, KaggleDBQA generated from Kaggle which include real time databases.

## 2.3 Natural language Question

The components of a natural language question include the main action or objective of the question, subjects in the question, qualities or characteristics of entities, constraints or filters applied to the query, operations that summarize data, and conditions that link multiple entities. Each component plays a crucial role in accurately translating the natural language question into a corresponding SQL query that can retrieve the intended information from a database. Two components of natural language question are a) Column and Table aware: system should be able to identify the column or table name from where user question can be fulfilled b) Text or main objective: this is to understand the user intension or what data user is trying to understand from the database system. Natural language questions and understanding user intentions present certain challenges. Based on survey from [7], user question is naturally ambiguous, which means that expressions can have multiple interpretations. Ambiguity can occur at various levels of natural language processing, such as lexical ambiguity, syntactic ambiguity, and semantic ambiguity. Lexical ambiguity occurs when a single word has different meanings, such as a word that can be a place name or a person's name. Syntactic ambiguity occurs when a sentence has multiple meanings. Semantic ambiguity refers to different semantic inferences. There can be context-dependent ambiguity in the question where the word has different meaning based on the question context. In such cases, paraphrasing can be challenging because different sentences can convey the same meaning but can be expressed differently. Other challenge is to understand missing information in a query, such as elliptical queries or follow-up questions, which requires inference. User mistakes, including spelling errors and syntactical or grammatical errors, can make the translation problem even challenging. As per [16]natural language challenges may include deciding how modifiers like adjectives and prepositional phrases relate to other elements in a question. One of the heavily researched areas under natural language question is understanding semantics of the questions. When a question or request is asked to a human resource [7], the research work would first understand the requirement by following; what is the requestor's request? where is the data stored? how to query the data? Going by this thought process, the system needs to be trained to understand and break down the process by this method. First understand the requirement or request from the user, then find out where the data is stored and finally, build the query. Understanding the user asks from the question is the first step for this system. The text to SQL conversion system can understand the user question only when it is converted to numerical representation. This is one of the key steps in this system. One of the most known techniques for converting questions into numerical representation is word embedding. With the development in NLP, like transformer [17] and pre-trained language model (PLM), there are better techniques for converting question into numerical representation. Word embedding is good at mapping each token from the user question into a unique numerical vector. One hot encoding is the simplest method of word embedding. GloVe [18] is a majorly used word embedding technique. [1, 11] used this method for word representation into numerical vectors. With the introduction of the transformer model [17] PLMs have become popular tools for NLP tasks. Similarly, PLM like BERT [19] are used for NL representation in text to SQL. There are two types of PLMs: Encoder only and Encoder-decoder models. Encoder based models take a sequence of words as input. The sequence starts with a special classification token (CLS), followed by the word tokens, and ends with a special separation token (SEP). Embedding module converts this sequence into an array of vectors representing these tokens. These are very robust when compared to Word embedding or GloVe. Encoder – decoder model like T5 [20], includes both encoding and decoding architecture within one system. These models take sequential words as input and generate sequences of text. They do not require additional layers to produce the output. Finding the database details like table name, column name and value entities is next step in the system. When a human resource tries to find the database information from a user question, they will review which part of the question mentions the database details. These parts of the questions can be classified as database information like table name or column name, query information like exact details or conditions to be queried. Connection between database and query can be called as schema information. When query information is mapped to a table or column, the value of a column is obtained. Finding schema information would be challenging for various reasons. Vocabulary used in the user question and query/database may not match. Users can request for information like actress name whereas database may store name of the actor under actor name. One other scenario would be when a user question references the value in a different way than how the data is stored in the database. Gender is saved as M or F for male or female. However, if the user question is looking for male employees in the organization, then the system should be trained to understand these value mapping to the database or query information. Schema linking has two types of tasks; one is to identify the query/database information, second is matching the query/database information. There are different ways to identify the query/database information. One of the ways is considering each word of the user question as query information. This may not be ideal as location or person name may not be always one word. Considering multiple words with the help of n-grams can be another way of identifying query information. [21]used the approach of n-gram to identify the table, query and value information from the user question. To identify the database entities, the author first groups the 1 to 6 lengths of n-grams from user questions. Then order them in descending order. If the n-gram matches with the database entities like table or column name, categorise them as table or column name accordingly. If n-gram is found to be in quotes, then it will be considered as a query value. Once n-grams are identified, overlapping n-grams are removed. Finally, all entities available in the question are identified and non-overlapping n-grams are obtained. [2]used an approach of n-gram length 2 to 6 to identify the column name from the user question. [22]used n-grams of 1 to 5 in question for determining the exact or partial match of the column/table name. [9] also used n-grams to identify the query values with more than one token. Identifying the place or person name using named entity recognition (NER) another technique in understanding the value condition in the user question. [9] used the NER model using transformer architecture. [23] tried to identify named entities like person, place, sports, company, and country using n-gram as keyword on Freebase. Another method to identify the database or value information is using knowledge base. [8]used knowledge graph ConceptNet to identify the values. To match the cell value with the column name from the database, the author first identifies the value in ConceptNet and connects it with the database. This system used search results with 'is a type of' and 'related term's types and assigned the column a type as exact match or partial match. Next step is matching identified information with the user question. After identifying the database/query information, making sure these are the right information is important. Identifying a technique to understand the semantic match is the next step. Exact or partial match is one simple method to find out matches. This approach was experimented by [8, 9] in the research work. If the table or column name exactly matches with the user questions words, then flag them as exact matches. If part of the word matches, then flag it as partial match. This also uses string matching with the database. For example, if the number exists in any of the database columns, then that is marked as a value match. [11] used an approach of 1 to 5 lengths of n-grams to determine exact match, partial match or no match. Once user intention, column or table information are identified, next step is predicting or building the query. This step includes how the input information is provided to the text to conversion system. The conversion system would require various input information like the user question, database details like table or column name etc. This information needs to be translated into the format in which the system can understand or process. Based on [24], the input can be provided to the system in four different ways. The early text to SQL systems like [1, 25] used the method of providing the user question and column information separately. These systems feed each word into bi-directional long short-term memory (Bi-LSTM) which will generate a hidden state for each word. Bi-LSTM also generates a final hidden layer for each column. At the end, both these hidden states are concatenated. Another approach can be providing all information from input into one sequence and encode them at once. PLMs like [19, 20] use this method. The input sequence in BERT is a concatenation of tokens from a single sentence or a pair of sentences, separated by a special token ([SEP]). Each token is represented by the sum of its token, segment, and position embedding. The segment embedding indicates whether a token belongs to sentence A or sentence B. The position embedding encodes the order of the tokens in the sequence. The first token of every sequence is always a special classification token ([CLS]). The final hidden state corresponding to this token is used as the aggregate sequence representation for classification tasks. [26] work uses a special token [CLS] to represent the whole input sequence and another token [SEP] to separate different segments in the sequence. The input sequence is the concatenation of question, column names and table names. The work uses the base version of BERT to encode the input sequence into hidden states. The hidden state of [CLS] is used as the context vector for decoding. The hidden states of column and table names are further processed by a BI-LSTM to obtain their representations. [4] forms one pair of input sentences for each column and concatenates a list of texts into one string. This is done by joining the texts with the blank spaces. Input sentence is further tokenized and encoded into base transformer model input. The token sequences are encoded to form a final input for the base transformer model. A graph is another way of representing the database information and their representation. [11] represents database schema as a directed graph and the nodes of the graph represent the columns and tables of the database schema. Each node is labelled with the words in its name. The edges of the graph are defined by the pre-existing relations within the database. These relations include primary and foreign key relationships. Additionally, the desired SQL program is represented as an abstract syntax tree (AST) in the context-free grammar of SQL. This representation is used to translate NL questions into SQL queries.

## 2.4 SQL Query

SQL follows syntax to execute a query. The query can be very simple with just a select statement or it can be complex with multiple joins and group by conditions. User questions can have spelling mistakes whereas SQL query needs to be syntactically correct to get the desired result. Some of the components of the SQL query are table name, column name and value mentioned in the query. The user question may not directly call out the column name, value or table name where the data is stored. This may pose a challenge in the way text is translated to SQL query. The data stored in different schemas may cause issues in the text to SQL translation.

## 2.5 Data Augmentation

Data augmentation for semantic parsing is one of the important techniques targeted to enhance the performance of natural language processing models, especially when training data is limited or lacks diversity. This technique involves creating data to expand and diversify the training dataset, thereby enhancing the model's ability to understand and generate precise database queries from user questions. Some of the works on this direction are listed below. The [11] presented paraphrases as a solution to the challenge of understand the user questions, building upon previous research in the field. This framework is distinctively trained using question-answer pairs, leveraging a neural scoring model that assigns weights to different linguistic expressions based on their likelihood of leading to correct answers. However, this framework does not have any reference about handling complex or ambiguous questions that require deeper understanding of context or background knowledge. Researchers in [8] focus on using a method called question generation to train a computer model with much less data than usual, like the WikiSQL dataset. The work highlights the importance of the logarithmic relationship between the volume of training data and the accuracy of semantic parsers, emphasizing the pivotal role of data quantity in boosting model performance. This research thus provides substantial insights into optimizing semantic parsers with limited supervised data, focusing on improvements in question generation and the effective use of training data. This work also highlights the limitations in question generation from SQL queries and effectively capturing details from the WHERE clause. To counter this, it advocates for an encoder-decoder model attuned to the structure of SQL, enhancing accuracy. Fundamentally, the study highlights the significance of incorporating both table information and external common knowledge, a strategy that aids in refining the parser's understanding of column types and inter-column relationships. In addressing the challenge of adapting semantic parsers to new environments without existing training data, [27]introduces Grounded Adaptation for Zero-shot Executable Semantic Parsing (GAZP). This innovative approach enables the adaptation of existing parsers by synthesising new, cycle-consistent data specific to the new environment. Distinguishing itself from traditional data-augmentation methods, GAZP generates consistency-verified data using a forward semantic parser and a backward utterance generator. Its compatibility with various model architectures across different domains, where cycle-consistency can be evaluated, makes GAZP a versatile and valuable tool in semantic parsing. However, this work leaves out several key challenges and areas for improvement. Firstly, it does not tackle the issue of handling out-of-vocabulary (OOV), or rare words present in input utterances. Secondly, this work does not address the complexities involved in managing complex or nested queries. There is also no discussion on how to deal with ambiguous queries where an utterance might lead to multiple valid SQL queries. The problem of processing noisy or incorrect input, which can significantly impact the accuracy and reliability of a semantic parser, is another aspect not covered in the work. [28]introduces a two-step data augmentation framework designed for cross-domain text-to-SQL parsing that operates without human intervention. The first step involves generating SQL queries based on an AST grammar. The second step is a hierarchical SQL-to-question generation model. This model breaks down SQL queries into individual clauses and translates each clause into a sub question. These sub questions are then concatenated to form a complete question, arranged in the order of the SQL query's execution. The effectiveness of this framework is validated through experiments on three cross-domain datasets: WikiSQL, Spider, and DuSQL. The results from these tests show that the framework significantly enhances performance compared to strong baseline models, with the hierarchical generation model playing a key role in this improvement. However, the work fails to address the challenge of varied natural language expressions for the same intent, crucial for parsing accuracy. It also doesn't tackle complex database schemas with multiple tables and relationships, which are common in real-world scenarios and require advanced encoding methods. Additionally, this work does not consider multi-turn interactions, such as follow-up questions or user feedback, which are essential for refining queries in a dynamic conversation context. The current methods for data synthesis in semantic parsing require manual rules, which restrict the exploration of diverse and unobserved data. [24] work proposes a generative model that features a probabilistic context-free grammar (PCFG) and a BART-based translation model that can efficiently learn from existing data and generate diverse unseen programs. The PCFG models the composition of programs (e.g., SQL), while the BART-based model maps a program to an utterance. By explicitly modelling compositions using PCFG, the generative model can explore unseen programs effectively, leading to the generation of diverse data. However, this work does not provide any details on improving the accuracy or efficiency of the semantic parser or parser optimization. Another method which was experimented in [10] is data balancing technique to handle the imbalanced dataset for nested queries in Spider dataset. The nested queries have different level of complexity like easy, medium, hard and extra hard. As per the research, while training the data, majority of the Text to SQL conversion systems handled these complexities as same and trained the model. This resulted in inadequate data representation of especially hard, extra query for the model and accuracy for the model for complex queries were low. To address these challenges, this research employed over sampling techniques to train the model. This model showed 5% increase in the accuracy when tested on spider dataset for hard and extra hard queries. In summary, data augmentation in semantic parsing involves a range of techniques like paraphrasing, data balancing, question generation, forward semantic parser, backward utterance generator, a hierarchical SQL-to-question generation model and a generative model for synthesizing the data.

2.6   Input Encoder

[29] introduced a method for sequence learning using Deep Neural Networks (DNNs). The authors chose this approach to overcome the limitation of DNNs in handling sequences of different lengths, common in problems like speech recognition and machine translation. This method uses a multi-layered Long Short-Term Memory (LSTM) model, which maps input sequences to a fixed-dimensional vector and then decodes the target sequence from this vector. However, one notable issue not completely solved is handling out-of-vocabulary words during translation. Also, while the LSTM approach effectively handles long sentences there is room for further optimization, particularly in terms of network architecture and vocabulary size. [1] uses the augmented pointer network designed to convert NL questions into SQL queries. The input sequence for the model is defined as the concatenation of column names, the question, and the SQL vocabulary. This way, the model does not need to generate tokens from a large vocabulary that may contain irrelevant words. The pointer network can directly copy words from the input sequence, which is useful for generating queries with rare words and column names that are not in the pre-trained word embedding. This approach is particularly effective for tasks involving sequential input and output, where traditional sequence-to-sequence models fall short, especially when the output categories vary based on the input size. However, this work only considers a subset of SQL queries that have a fixed format and a limited vocabulary.

It does not handle complex queries that involve joins, sub queries, nested aggregations, etc. This work uses a pointer network to generate the SQL query from the input sequence, which may not be optimal for capturing the compositional and hierarchical nature of SQL syntax. In [2] each word in the user question is tokenized using n-grams. The system uses the tokens to search in the table schema and if they are found, are renamed as column name. Also, identifies the numbers and named entities in the question. This work uses two bi-directional LSTM networks to separately encode words, types in the question and the column names separately. By analysing the types of entities and numbers mentioned in the questions, TypeSQL can accurately interpret the users intent, leading to precise query generation. This approach is particularly effective in handling questions involving specific entities or numerical data, where conventional methods might struggle. Although TypeSQL utilizes type information to better understand rare entities and numbers in the questions, it still faces challenges in handling queries that do not have any column name and entity indicator. This makes generating the right SQLs without searching the database content impossible in such cases.

## 2.7 Output Decoder

Based on [30] the encoder-decoder model can be categorized into three broad categories. Sequence to sequence approaches, grammar-based approach and sketch-based slot filling approach. Sequence based decoder generates the SQL query as a sequence of words. This was used by [1] model which was the first model to introduce deep learning text to SQL systems. One of the drawbacks of this approach was grammatically incorrect SQL queries. However, these models are gaining attention with the introduction of pre-trained transformer models. [31]is one of the examples. Sketch based slot filing decoder approach involves predicting certain parts of the query. The model will have a query sketch and should fill the slots with the predicted values. [25] used a sketch which is like SQL grammar and the model will fill the slots with the predicted output/values. Sketch captures the dependency of the slots to make predictions. [23] uses a similar approach with the additional feature of knowledge-based type awareness. Since this includes filling the slots in the sketch, the implementation of this approach is complex and difficult to scale it for complex queries. Grammar based decoder models produce a sequence of grammar rules as their output. These are the instructions for creating a structure query. [30]uses a sketch-based slot filling decoder to fill the slots of each sketch. [11]uses a tree-structured decoder that generates the SQL query as an AST in depth-first traversal order. It uses an LSTM to output a sequence of decoder actions that either expand the last generated node into a grammar rule or choose a column/table from the schema. Decoder system in [8]is a grammar-based neural model that synthesizes SemQL queries from NL questions. The decoder interacts with three types of actions: APPLYRULE, SELECTCOLUMN, and SELECTABLE. The decoder uses a LSTM to model the generation process of a SemQL query via sequential applications of actions. The decoder also employs a memory augmented pointer network to select columns from the schema or the memory.

## 2.8 Evaluation

After predicting the SQL query for the user question, the system should be able to validate if the predicting query is accurate. The dataset should have the gold source or ground truth query for each user question. That will enable the system to test and evaluate its results. Different types of metrics used for evaluation of the text t SQL systems are listed below. Logical form accuracy: This metric evaluates if the predicted SQL query matches with the gold source query available with the dataset. The output is marked as correct only when the predicted query is written in the same way as it is in gold source. [1]used this method to validate its output. Execution accuracy: In this metric, gold source and predicted query is executed against the dataset. If the result is the same, then output is marked as accurate. These [1, 23]research work used this method to validate output. Component matching: This was introduced by [23] to understand which part of the SQL query is accurately predicted by the system. In this metric, each component is decomposed in predicted and ground truth query and evaluates if these two components exactly match. SQL Hardness Criteria: [23]introduced this metric where SQL queries were divided into 4 levels: easy, medium, hard and extra hard based on the number SQL components. Model performance in this research [10] was measured using exact matching accuracy on different levels separately.

## 2.9 Different Systems

[1] introduced a deep neural network for user questions into SQL queries. The model leverages rewards from in-the-loop query execution over a database to learn a policy for generating the query, improving execution accuracy and logical form accuracy. Network classifies the aggregation operation first and then moves to the column name in the input data to the select section, finally generating where condition using the pointer network. The work used execution accuracy to measure the percentage of queries that, when executed, produced the correct result. Logical form accuracy to measure the percentage of generated queries having an exact string match with the ground truth query. The authors introduce WikiSQL, a large dataset of hand-annotated examples of questions and SQL queries from Wikipedia, which is used to train and evaluate the Seq2SQL model. One of the limitations of the Seq2SQL model, as discussed in this work, is related to the use of cross entropy loss for network optimization. The issue arises when WHERE conditions in a SQL query can be swapped without affecting the query's result. To address this, this work applies reinforcement learning to learn a policy that directly optimizes the expected correctness of the execution result, rather than relying solely on teacher forcing at each step of query generation. However, the accuracy of this network using reinforcement learning (RL) increased only by 2% (from 57.1% without RL to 59.4%). [25]highlights that existing approaches using sequence-to-sequence models encounter an issue known as the "order-matters" problem, particularly when the order of elements is irrelevant. The proposed sketch-based method employs a sequence-to-set and column attention mechanism without relying on reinforcement learning. The sketch-based approach utilises dependency graphs within the sketch to make predictions based on the previous prediction it depends upon. SQLNet predicts the value for each slot in the sketch. The sketch model is intentionally crafted to be generic, allowing it to accommodate all types of SQL queries. This design decision effectively addresses the challenge of "order matters" faced by sequence-to-sequence models. The column attention is used to identify the relevant columns in the query synthesis process. SQLNet outperforms the state-of-the-art approach, Seq2SQL, by 9% to 13% on the WikiSQL task. [23]designed SyntaxSQLNet to tackle the intricate and multi domain task of text to SQL generation and evaluated on the Spider dataset. The author uses a database partitioning strategy in which the database is divided into separate training and test sets. This model breaks down the SQL decoding process into 9 different steps for predicting different SQL components. It organizes the decoder by considering SQL grammar, which determines the appropriate module to invoke at each decoding step. The input encoder utilises a bidirectional LSTM to encode a question sentence, incorporating table and column heading to construct the column embedding. Moreover, the system leverages the current decoding history of the SQL query as input for each module, using previous states to predict the subsequent SQL token. The system adopts a sketch-based approach for each module inspired by [25]and additionally, it leverages the seq2set prediction framework. The [11] aims to tackle the generalisation challenge by (a) encoding the database relations in a format that is accessible for the semantic parser, (b) modelling the alignment between database columns and their corresponding mentions in each query. This work introduces a framework for a text to SQL encoder that handles schema encoding, schema linking, and feature representation by leveraging the

relation-aware self-attention mechanism. According to the authors, this representation learning can prove advantageous for tasks beyond text to SQL, if the input exhibits some predefined structure. Based on the error analysis performed on the system, 39% of the errors are due to a wrong, irrelevant or missing column in the select section. This is the limitation of the schema linking method due to unambiguous references. 29% of the errors are due to missing a where clause. One of the major reasons for this error is lack of domain-specific phrasing like older than 21 which requires the system to understand to map the older than 21 to age column. These errors may require domain specific fine tuning. When large pre-trained language models are fine-tuned for constrained formal languages like SQL, they frequently produce invalid code, rendering it unusable. [31]proposes a method called PICARD that constrains auto-regressive decoders of language models through incremental parsing. PICARD verifies the output at a grammatical level. It can reject invalid query structures and detect issues with compositions of SQL expressions. This method seamlessly integrates with existing algorithms for greedy and beam search, commonly employed in auto-regressive decoding from language models. This approach limits predictions to the top-k highest probability tokens and assigns a negative score to those that do not pass PICARD's checks. PICARD offers four mode settings, a) no checking, b) lexical analysis, c) parsing occurs without additional safeguards, d) parsing includes additional protective measures. The experiments primarily concentrated on the Spider dataset. [8]this research introduces an approach called IRNet that addresses the mismatch between intents expressed in spoken language, the specifics of SQL's implementation and the difficulty in foreseeing columns that result from words outside of domain. IRNet breaks down the entire process into three phases: schema linking, SemQL query synthesis, and deterministic inference of SQL queries using domain knowledge. This work highlights the effectiveness of using intermediate representations in various semantic parsing tasks and the success of IRNet on the Spider benchmark, where it achieves the first position on the leaderboard. There are efforts to improve column prediction when the column name is missing in the question or when synonyms of the column are used, especially when values exist in the question. It also mentions handling operators like ASC (ascending) or DESC (descending) by parsing phrases such as "old to young" to order the data accordingly. This work also highlights limitations of SemQL which lacks support for self-joins and does not entirely eradicate the disparity between NLQ and SQL. The use of IRNet with BERT (a language representation model) has shown to provide better accuracy than using IRNet without BERT. [26]presents a new neural architecture for Text-to-SQL over single tables by utilizing column contents, which has not been fully exploited before. This work introduces a representation learning method that enhances the representation of columns and queries for Text-to-SQL tasks. This work addresses the challenges of extracting values accurately from different columns and inferring the right values when column names are not explicitly mentioned in the query. This approach involves multiple layers in the neural network, including input encoding and, the use of BERT for improved column representation, an attention mechanism for better question representation, and multi-task prediction for slot-filling in the generated SQL. Using pre-trained BERT models, it improves feature extraction from columns and query reformulation. The model shows better accuracy on benchmarks like WikiSQL and a Chinese dataset, TableQA, especially in column prediction and handling semantic ambiguities. The approach signifies progress in semantic parsing, particularly in constructing SQL queries from natural language inputs. This work also did a manual error analysis on a set of failed predictions. One of the identified problems was the value extraction problem, where the selection of an erroneous column led to mistakes in the SQL statement generation. Even after introducing column content, cannot be accurately interpreted by the model. This work suggests that external information might be necessary to improve accuracy. This work suggests two future enhancements to ER-SQL model. Firstly, the current limitation of BERT model and system crashes due to long operation time on a large dataset. This can be addressed through optimization of efficiency and integrating a Word2Vector approach on a larger corpus for better performance. Secondly, improve the similarity calculation. The model uses the Rouge-L algorithm to calculate the similarity between the question and the table content. These can be improvised by using graph applications. [9] is a research work that introduces two systems ValueNet light and ValueNet. These systems are designed to convert natural language questions into SQL by incorporating values extracted from the database content and metadata. This work highlights the challenge of incorporating values into SQL queries, which is often overlooked by previous systems. ValueNet light and ValueNet use novel pre-processing techniques and neural network architectures to achieve this. This work extends the [8] framework with transformer architecture to encode values extracted from a database, addressing the challenge of value representation in complex SQL queries. The system includes pre-processing for identifying schema elements, an abstract syntax tree (AST) for resolving mismatches, and a decoder using LSTM and pointer networks. ValueNet Light and ValueNet are presented as solutions to incorporate explicit and implicit values in queries, with ValueNet providing an end-to-end approach that includes value extraction, candidate generation, validation, and encoding, utilizing NER models, heuristics, and Damerau–Levenshtein similarity. This work uses the Spider execution accuracy metric to evaluate the performance of ValueNet. This metric includes the proper prediction of values, unlike the exact matching accuracy metric used in previous systems. Based on the error analysis presented in this work, 50% of the errors are due to VaueNet failing to predict the correct column. This is majorly due to similar names across different tables which is difficult to identify. 39% of the errors were due to SQL sketch errors which might be due to lack of domain specific information. One of the future works mentioned in this work based on error analysis involves improving the architecture sketch for generating good value candidates. This could include research into generative neural network approaches, such as those based on text Generative Adversarial Networks (GANs) and leveraging available data from the database to enhance the system's ability to come up with accurate value candidates. [10]this work seeks to enhance the precision of translating NLQ to nested SQL queries. This method refines the IRNet framework by incorporating data oversampling, incorporates database content and creates a new version of the SemQL grammar and updated loss function. Data oversampling is used to boost the resemblance of complex queries, while the novel loss function accounts for the complexity of SQL queries. Looking at the type of error in IRNet [8]the failure corresponds to complex queries and this work considers the problem could be due to an imbalanced dataset. This work also emphasises adjusting the loss function as a suitable method for imbalanced data. To evaluate the method, the author employs component matching and SQL hardness criteria. This work also mentions advanced embedding techniques used to improve the accuracy of text to SQL translation. This work suggests future work could involve applying these methods to other model architectures and further refining the loss function to boost overall accuracy while specifically improving nested query predictions. [32] this work proposes a framework that decouples schema linking and skeleton parsing in Text to SQL. The seq2seq model is responsible for parsing both the schema items (tables and columns) and the skeleton (SQL keywords) of SQL queries due to their structural properties. This makes it difficult to parse the correct SQL queries, especially when they involve many schema items and logic operators. This work proposes a ranking-enhanced encoding and skeleton-aware decoding framework to decouple the schema linking and the skeleton parsing. The framework injects the most relevant schema items into the encoder, which alleviates the schema linking effort during parsing. The decoder produces the SQL slot and then the actual SQL output, inherently constraining the SQL analysing process. It shows that the column-enhanced layer in the encoder helps alleviate the table missing problem, and the focal loss in schema item classification is effective in handling label imbalance in the training data. [33]this work introduces a new Task decomposition,

Knowledge acquisition, and Knowledge composition (TKK) framework that aims to boost the adaptability and resilience of text to SQL parsing by breaking down the process into task decomposition, knowledge acquisition, and knowledge composition stages. In practice, text-to-SQL parsers often encounter various challenging scenarios, requiring them to be generalizable and robust. The TKK framework tries to solve this challenge by enabling the parser to gain comprehensive SQL knowledge rather than capturing irrelevant patterns. TKK features a prompt-based learning strategy to separately acquire the knowledge of subtasks and employ the learned knowledge to tackle the main task, i.e., generating the entire SQL query. In the knowledge acquisition stage, TKK trains the model with all the subtasks in a multi-task learning manner; in the knowledge composition stage, TKK fine-tunes the model with the main task to combine the acquired knowledge of subtasks and learn the dependency between them. In diverse scenarios, the TKK framework demonstrates effectiveness and achieves state-of-the-art performance on the Spider, SParC, and CoSQL datasets. This work demonstrates the effectiveness of the TKK framework through comprehensive evaluation on three levels of generalisation (i.i.d., zero-shot, and compositional) and robustness. The TKK framework outperforms previous models such as T5 and shows strong zero-shot and compositional generalisation abilities. Although the TKK framework is conceptually simple, it needs to decompose the task into multiple subtasks manually. Therefore, a general strategy to automatically discover the meaningful substructure of the original task is needed. [34] introduces a "Hybrid Decoder," which combines generation-based and sketch-based methods, offering a significant advancement in accurately and efficiently generating SQL queries. The author studies how to leverage pre-trained language models in Text-to-SQL. As per author the previous approaches underutilize the base language models by concatenating all columns together with the NL question and feeding them into the base language model in the encoding stage. This approach breaks down the problem into column-wise ranking and decoding. Finally assembles the column-wise outputs into a SQL query by straightforward rules. The encoder is given a NL question and one individual column, which perfectly aligns with the original tasks BERT or RoBERTa is trained on, and hence the system avoids any additional encoding layers which were necessary in prior approaches. Key feature is the Value Prediction Module, which simplifies decoding and reduces vocabulary size, enhancing the model's performance. The model showcases superior performance in mutual information reflection and syntactic accuracy, along with faster inference speeds compared to existing methods. [35] addresses the challenge of Text-to-SQL translation with limited training data. It proposes a divide-and-conquer framework called SC-Prompt, which splits the Text-to-SQL translation into two simpler sub-tasks: a structure stage and a content stage. The structure stage focuses on generating SQL structure with placeholders, while the content stage fills these placeholders with specific values. The study utilizes T5 as the base pre-trained language model and introduces a hybrid prompt strategy with learnable and fixed vectors, aiming to guide the PLM effectively for better context understanding. Additionally, this work employs keyword constrained and structure-guided decoding for better SQL validity and accuracy. The model's success is evaluated through exact set match and execution accuracy metrics. The approach is shown to better performance in few-shot scenarios, particularly on the Spider dataset with limited training examples. In the future, the author assumes the SC-Prompt framework can be applied for multiple semantic parsing tasks. Also, use transformer-based models with a limited input length (e.g., 512) to encode extra-large databases. [36]explores the use of Large Language Models (LLMs) for Text-to-SQL conversion in the power grid domain, a scenario where domain-specific data is scarce. A specific prompting method with ChatGPT is used for efficient data generation, improving annotation. The study introduces a pre-trained electricity domain LLM with instruction tuning for SQL generation. The approach involves a two-step task division: schema prediction and SQL generation, using models like XLM-RoBERTa and a transformer-based model incorporating domain knowledge. The study measures success using exact matching accuracy and execution accuracy, acknowledging the difficulty in handling complex queries and the generation of non-executable SQL statements. Future directions include using different architectures; targeted instruction-based fine-tuning to effectively enhance Text-to-SQL capabilities. [37] presents a novel interface for querying network performance using natural language, specifically designed for Wireless Mesh Networks (WMN) with non-semantic database nomenclature. It overcomes challenges of incorrect transcriptions and noisy inputs via state-of-the-art model fine-tuning, semantic view creation, and domain-specific text corrections. System includes a) semantic views creation in the database and dataset update: this algorithm involves replacing non-semantic names in the SQL queries with new semantic table and column names, b) domain-specific corrections performed on natural language text inputs; this algorithm involves applying regular expressions to correct potentially wrongly transcribed entities, such as IP or MAC addresses, and correcting any incorrect variations of categorical values in the database columns, c) fine-tuning Semi-autoregressive Bottom-up Semantic Parsing (SmBoP) translation model, with networking domain-specific data to enable text-to-SQL translation for a WMN. The interface shows high exact match and execution accuracy. Future work can involve integrating intent-based and context-aware querying for complex questions with this system. Query-based split of the dataset for fine-tuning may also be considered as a future work. [38] discusses a time-efficient SQL query generator for cross-domain datasets, using a user's interaction history to enhance query generation quality. It employs encoder-decoder architecture with a query attention mechanism that capitalizes on the linguistic dependencies between successive queries. This methodology allows for minor modifications to prior queries, making it suitable for real-life applications. This literature's proposed approach includes a bi-LSTM and BERT-based sentence-table encoder, a unidirectional LSTM interactivity encoder, and a table-aware decoder with query editing capabilities. Sentence table encodes the sentence tokens at each turn. The model uses two kinds of inputs for the bi-LSTM: word weights from a pre-trained model and contextualized BERT word weights. This step groups all the heads of the columns into a single sequence separated by the [S] token. This encoder captures information across the sentences given by the user and is positioned above the sentence-level encoder. It uses the last stage of hidden layer phrase encoding as input to a unidirectional LSTM interactivity encoder. Table decoder feeds the history from the interactivity encoder, the current sentence, and the table schema into an LSTM decoder with observation to construct SQL queries. Query editing mechanism is added to the table decoder to handle closely related queries that frequently overlap with previous ones. It involves copy and inserts operations on the prior query, using another bi-LSTM model to encode the previous query. The context vector is extended with attention to the previous query, and a switch is predicted at each decoding step to determine whether to copy from the previous query or insert a new token. The system includes three different algorithms: encode algorithm, attention mechanism, decoder algorithm. This work assesses performance using component matching, specific match, and execution accuracy metrics. Future work may involve creating a comprehensive data collection framework for better model training. [39]this work discusses enhancing encoder-decoder models for text-to-SQL conversion by integrating higher-order semantics and structured knowledge of SQL into a unified framework called HG2AST. It introduces an encoder with dual graph message passing and a grammar-based decoder that eliminates permutation biases through a Golden Tree-oriented Learning (GTL) algorithm. This work introduces four basic modules, two for the encoder and two for the decoder. This work uses attention-based pooling function and Relational Graph Attention Network (RGAT), the transformer architecture can be regarded as a specific implementation of Graph Attention Network (GAT), in the encoder. Multi-Head Cross-Attention (MHCA) and pointer network, it is widely used in the decoder to copy raw tokens from the input memory, in the decoder.

HG2AST framework is split into four parts: 1) a contextual input module, 2) a structural encoding module, 3) two auxiliary modules, and 4) an AST output module. The contextual input module initializes features for all nodes and edges. PLM and three type-aware bidirectional LSTMs are used here. Next, the structural encoding module considers the structures of both the node-centric and edge-centric graphs, and updates features of nodes and edges layer wise. After getting encoded node features X, author appended two auxiliary modules, namely value recognition and graph pruning, to construct the value/table/column memories V/T/C for the AST decoder to retrieve or copy from the auxiliary modules. The approach advances the field by refining the interaction between heterogeneous graph neural networks and abstract syntax tree construction, improving the generation of SQL queries from natural language. This work also suggests that future research could benefit from sophisticated graph structure knowledge and advanced techniques in entity linking and value normalization. [40] presents an approach to improve text-to-SQL translation models that suffer from limitations on diverse datasets and complex queries. This work extends the SQLSketch to make the model aware of the data types of columns. Second, proposed a compatibility-based approach that allows the prevention of the generation of faulty queries that can't be executed against database engines. Third, this work provided a new approach to better make predictions of values from the input NL sentences. To make the model aware of the columns' types', the data types are included in a new way in the structure of the input. In the second part, the reduced schema of the database is fed after predicting the number of tables and the candidate ones from the question using hand engineering methods such as Levenshtein Distance. The table name is added as a prefix for each column. The NL sentence is separated from the columns by the special token [SEP] which also separates the elements of each column. A type-compatibility based approach checks the agreement of items within the target SQL query and removes elements from the list of candidates' items in each clause. This prevents the generation of faulty and wrong queries that will trigger errors when executed against the database engine. This work validates the incompatible items in SELECT, WHERE, HAVING and ORDER clauses before the final prediction. A combination of regular expressions which can return numeric, dates, times, Booleans and string literal values directly from the questions from the GreatSQL dataset clauses are used to extract all the values from the NL sentence. The approach is evaluated using component matching scores and exact matching accuracy. Future work may focus on translating right to left languages, handling aliases, and correlated queries, as well as including execution matching as a metric. The model's reliance on hand-engineered techniques for predicting tables and clauses may need refinement for better scalability and accuracy across various domains and schemas. [6] presents a novel approach to enhance text-to-SQL generation using transformer-based sequence-to-sequence models. The authors introduce Schema-aware Denoising (SeaD), which involves two denoising objectives that help the model recover input or predict output from specifically designed 'erosion' and 'shuffle' noises. These denoising tasks serve as auxiliary tasks, improving the modelling of structured data in sequence-to-sequence generation. Additionally, this work proposes a clause-sensitive execution-guided decoding strategy to address limitations in existing execution-guided decoding methods. The experimental results demonstrate that this method significantly improves the performance of text-to-SQL models in terms of schema linking and grammar correctness, setting a new state-of-the-art on the WikiSQL benchmark. This work [5]presents a novel based on RoBERTa to convert natural language questions into SQL queries while prioritizing data privacy. This algorithm is designed to generate SQL queries from natural language questions without relying on table data, focusing instead on the table schema. The authors utilize RoBERTa embeddings and data-agnostic knowledge vectors in LSTM-based sub models for this conversion. The natural language questions and table schemas were tokenized using RoBERTa's default tokenizer. This process involves breaking down the text into smaller units (tokens) and converting them into IDs from the model's vocabulary. System includes generation of two types of knowledge vectors, the question mark vector and the header mark vector. These vectors are Boolean and encode the significance of headers and tokens in the question for the model, aiding in the prediction of SQL queries. These pre-processing steps are critical for transforming the input data into a format suitable for model processing and for capturing the relational information between the natural language questions and the table schema. Their method excludes table data during training, aiming for zero-shot learning capabilities and ensuring data privacy. This approach aims to address the challenges of data privacy and the need for efficient natural language to SQL query generation. Model achieves 76.7% execution accuracy on a test set, demonstrating its effectiveness while maintaining data privacy. This work also discusses future enhancements, extending this system to multiple table queries; support complex SQL queries and clauses. [4] introduces HydraNet, a novel approach leveraging pre-trained language models like BERT and RoBERTa for text-to-SQL tasks. HydraNet focuses on column-wise ranking and decoding, which aligns with the training tasks of base language models. This method avoids additional pooling or complex layers, simplifying model structure and enhancing efficiency. This work demonstrates HydraNet's effectiveness on the WikiSQL dataset, achieving top results. The approach addresses major challenges in text-to-SQL conversion, particularly in fusing information from natural language questions and table schemas, ensuring output SQL query executability and accuracy, and effectively utilizing pre-trained language models. As a future work, this work recommends expanding this system to full SQL grammar. [41] focuses on developing an application that converts natural language questions into SQL queries for use in e-learning contexts. It addresses the challenge of SQL complexity, aiming to provide real-time database access without requiring SQL expertise. The methodology involves using datasets from the e-Yantra robotics competition and training models like [11]and SmBoP on these datasets. The model involves the SmBoP architecture for training the Text-to-SQL task. Additionally, a RAT-SQL encoder is employed to encode the input and schema. This encoder comprises 24 transformer layers followed by 8 RAT-SQL layers. The decoder used in this setup is SmBoP.  SmBoP is a bottom-up parsing algorithm that builds the top-K sub trees of height less than T at each decoding step T. The algorithm constructs these top-K program sub trees in parallel at each decoding step, a process like beam search. This approach results in a runtime complexity that is logarithmic in tree size, rather than linear, making it an efficient solution for AI systems. The algorithm also utilizes cross-attention to contextualize the trees with information from the input question. Results show promising accuracy rates. In addition to these quantitative measures, the study also involved qualitative evaluation through detailed analysis and testing by the office staff at e-Yantra. The feedback from this test group indicated that it was significantly easier to obtain simple data in the form of tables and graphs by using the system to interpret and process data requests formulated in natural language sentences. Moreover, there was a notable reduction in the time taken for the staff to access and analyse the data, highlighting the practical benefits and efficiency of the system developed in the study. For future work, this work suggests exploring generalizability across different domains and improving accuracy.

2.10   Summary

Based on the above review, it can be understood that there are many systems developed for converting natural language questions to SQL. The systems used different mechanisms to understand the user question and help the system to translate it accurately or predict executable SQL query. Even then, there are issues with predicting the correct SQL or executable SQL. Out of many reasons, missing where clause specially when the questions include domain specific phrases like smaller than, less than or bigger than etc. The size of the data with these domain specific queries are limited and training the model on these query pattern would be difficult without balancing the data. To solve this issue, this research work would use data balancing technique to over sample these data set and fine tune the model training on subset of domain specific data. These two methods would help in increasing the model performance on the natural language question to SQL translation.

## 3. RESEARCH METHODOLOGY

This work focuses on improving the accuracy of the text to SQL conversion. The model should be able to provide accurate information for the natural language questions from the user. This work focuses on data balancing technique for domain specific data, modifying the architecture to extract good value candidates from the user question. This work utilizes WikiSQL for the domain specific data balancing experimentation. This is a dataset of 80,654 hand-annotated examples of questions and SQL queries distributed across 24,241 tables extracted from Wikipedia. The dataset comprises collected questions, their corresponding SQL queries, and associated SQL tables. Specifically, 56,355 samples serve as training data, 8,421 for development, and 15,878 for evaluation. Each natural language question in this dataset can correspond to multiple SQL queries. These SQL statements typically include a SELECT clause with a maximum of one aggregate operator and a WHERE clause with a maximum of four conditions joined by the AND operator. Some of the pre-processing techniques that may be needed for NL question as part of this research are.

- Removing stop words
- Stemming and lemmatization to normalise NL question to remove the randomness of a text and make it consistent with a predefined format
- Breaking down the words in the NL question to token
- Part of Speech (POS) tagging
- Named Entity Recognition (NER) to identify and classify the entities to place, name, location or organisation
- Parsing to identify the grammatical structure to understand the meaning of the question

3.1     Data Augmentation

Based on [10] one of the methods which can be used to treat the data is data balancing technique. Especially when the distribution of queries is uneven in the dataset data imbalance can significantly affect the model's performance. When the data is underrepresented compared to other observations, the model may lack generalizability for the unseen data. Since model is not trained on the underrepresented observation during training process, system will fail to make accurate predictions for these examples. This will result in missing where clauses in the generated SQL from the system. To overcome this challenge, data balancing technique can be utilized. Two major techniques data balancing technique are discussed below: Under sampling: Decrease the number of majority observations by randomly removing them from the dataset with or without replacement. One of the disadvantages of this method is it may remove some useful observations from the dataset. Hence this method will not be considered for this research work. Oversampling: Increase the number of observations for underrepresented queries by duplicating existing queries. This method can be used when there are no sufficient observations for training the model. The process involves the below steps. To balance the dataset, calculate the number of additional observations needed for the underrepresented observations. To achieve the balanced data set, create new observations of the underrepresented data by sampling with replacement from underrepresented dataset. Sampling with replacement means each sample is chosen multiple times from the dataset. Here is how it works. First define the dataset.

Dataset with normal query is referenced as majority observation and domain specific query as minority observation. Next calculate the number of additional samples needed to balance the date. The difference between number of majority samples and minority samples are considered as additional samples required to balance the dataset. Post that, create oversampling dataset. Oversampled dataset created by duplicating data records for the minority samples.

*majority observation Dataset ($D_{majority}$): contains $n_{majority}$ samples*

*minority observation Dataset ($D_{monority}$): contains $n_{minority}$ samples*

$$m = n_{majority} - n_{minority}$$

where: m is additional records required to balance the underrepresented data to the same size as majority data.

$D_{oversampled} = D_{minority} \cup \{x \mid x \in D_{minority}, x \text{ is the sampled with replacement}, 1 \leq I \geq m\}$

where: $D_{oversampled}$ is the new dataset that combines the underrepresented observation with duplicated data records.

x represents each new sample from underrepresented dataset i is an index used to iterate through the number of additional samples (m) needed for the balanced dataset.

To understand this better, here is an example for the above:

i. $n_{majority}$ = 300 records
ii. $n_{minority}$ = 100 records
iii. Additional samples needed m = $n_{majority}$ − $n_{minority}$ = 300 − 100 = 200 records
iv. Oversample dataset $D_{oversampled}$ = 100 (original minority records) + 200 (additional samples) = 300 records.

This results in balanced dataset since both majority and oversampled dataset includes 300 each record.

### 3.2 Modelling Techniques

Based on the taxonomy from survey review [7] Text to SQL conversion includes the following systems. Schema linking: When humans create SQL queries, they naturally associate NLQ components with relevant database elements. Similarly, text to SQL system could gain advantages by employing a similar approach during NLQ translation. Natural Language: One of the key parts of the text to SQL model is processing the natural language questions. The model can understand only numerical values. The NL questions need to be converted into numerical representation. Word embedding is one of the methods used for this process till the transformer models introduced for NLP. With the introduction of transformers and their use in PLMs have been made great improvement in NLP. Few of the key features of PLMs are they are trained on large datasets, they can be fine-tuned for NLP specific tasks like text classification, question answering, summarization and transfer learning. PLMs such as BERT and RoBERTa, function as encoder-only models. They process input sequences and generate contextualised numerical representations for each token. Unlike word embedding techniques that map individual words to fixed vectors, PLMs compute representations using all input tokens, capturing the context and meaning of words and phrases rather than isolated representations. Encode-Decoder models, such as T5, are comprehensive end-to-end architectures tailored for handling sequential text inputs and producing corresponding outputs. These models autonomously generate the output without any additional neural network. Their versatility extends to various downstream tasks.

### 3.3     Input encoding:

The input encoding dimension relates to the structuring and feeding of input into the system's neural encoder for efficient processing. The NLQ along with the database (DB) column and table names inputs are necessary. Neural networks necessitate converting all inputs into a format compatible with the network. Some of the encoding methods are, encode NL questions separate from table column. Example for this method is [1, 25]. Input serialisation is another approach involves serialising all inputs into a single sequence and encoding them simultaneously. This practice is quite common with PLMs like BERT [19] and T5 [20], which generates a contextualised representation of input. Encoding each input individually would hinder the system from fully utilising the contextualization capabilities of PLMs. Instead, this method streamlines the encoding process and leverages the inherent robustness of PLMs. Encoding each column separately is another method of input encoding. [4]employs a distinctive strategy. This analyses the question for each column separately and generates results for each column autonomously. For each table column, a unique input is created by combining the NLQ with the column name, type, and table name. Schema graph encoding represents database elements and their relationships using a graph are highly effective. [11]is one such example.

### 3.4     Output decoding:

The decoder architecture can be categorised into three types according to their decoder's output generation method and they are as below: Sequence-based: This was used by the first deep-learning model for text to SQL conversion  [1] and this is the simplest method. The system produces the predicted SQL query as a sequence of words, including SQL tokens and schema elements. The limitation of this approach lies in its tendency to produce inaccurate queries, as it deviates from the standard SQL structure. However, with the introduction of PLMs like BERT [19], T5 [20], this method is used again and these PLMs prevent from generating incorrect queries.  Grammar-based: These are advancement from sequence-to-sequence methods. Instead of generating a sequence of basic tokens, they output a series of grammar rules. These rules help in constructing a structured query. E.g.: [8, 11] etc.  Sketch-based slot-filling strategies: This simplifies the task of predicting SQL query by predicting various parts of the query This way the prediction problem is transformed into classification problem. Various models like [4, 23] uses this method

### 3.5     Neural Training:

This is the methodology used to train the text to SQL conversion system. There are three ways to train the system. Fresh Start is the common method to train the system from scratch. Transfer learning is second method. With the introduction of transformers and PLMs, this method is gaining popularity. It entails leveraging a model initially trained on a broader task and dataset, and then incorporating it into a new model that is subsequently fine-tuned for a specific downstream task, such as text to SQL parsing. This approach allows the downstream model to benefit from the pre-existing knowledge captured by the pre-trained model, enhancing its performance on the specific task. Additional objective is the third method. This approach aims to enhance the model's performance by leveraging diverse learning objectives and pre-training strategies. Some of the ways to do this are erosion, shuffling. graph pruning. Schema dependency learning, Pre-training specific components involves training distinct parts of network to adapt effectively to the unique characteristics of a given task.

## 3.6 Output refinement

This approach can be applied to a trained model to achieve enhanced results or prevent the generation of incorrect SQL queries. This is used with the [1] model. One of the techniques for output refinement is execution-guided decoding. This strategy avoids generating SQL queries that cause execution errors.

## 3.7 Algorithm and Techniques:

State of the Art Models: Below are the recent pre-trained texts to SQL conversion systems.

- SQLSketch-TVC
- SQLSKetch
- HydraNet
- IRNet
- Seq2SQL
- SQNet
- RAT-SQL
- HG2AST framework
- Text2SQLNet
- RYANSQL
- LGESQL
- SyntaxSQLNET
- CRUSH4SQL
- SeaD

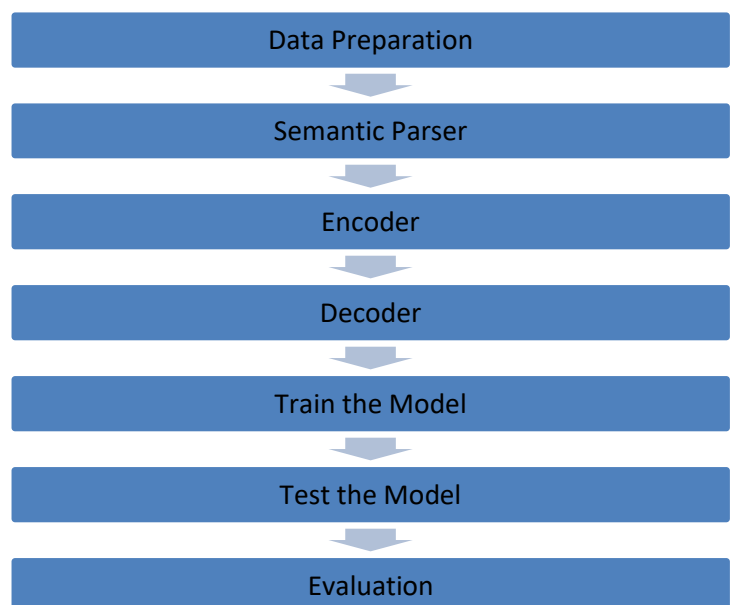

Figure 1 3.10 Text to SQL Conversion System process flow

Figure 3.10 provides High-level research methodology end to end process

## 3.8 Proposed Method

This work would be experimenting on refining [11] methodology to improve the value conditions generation through data balancing technique on domain specific questions. In addition to this, fine tuning the model to increase model's ability to understand domain specific question.

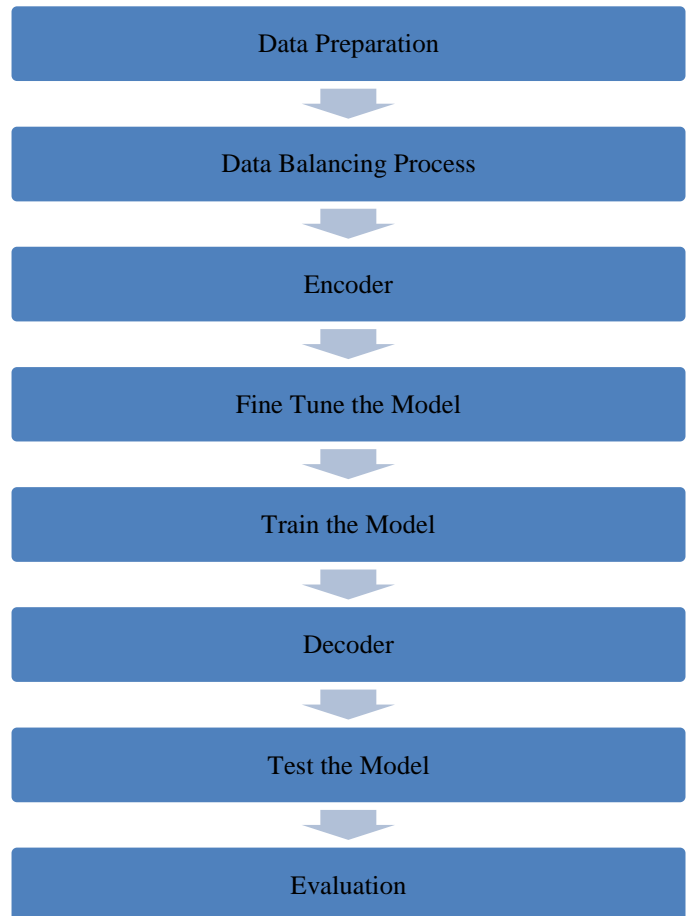

Figure 2 3.11 Proposed Solution process flow

Figure 3.11 shows the proposed methodology experimented in this thesis.

## 3.9 Evaluation Metrics

This work will be using following metrics to evaluate the performance of the proposed model.

Logical form accuracy: This metric quantifies the number of queries that exactly match the ground truth query used during paraphrase collection. It assesses how closely the generated queries align with the reference queries. Achieving high logical form accuracy is crucial for ensuring that the generated SQL queries correctly capture the user's intent and adhere to the expected structure. Therefore, this work will utilise this metric to gauge the effectiveness of my approach.

*Logical Form Accuracy = Number of Correct SQL queries ÷ Total Number of SQL queries*

Execution accuracy: This quantifies the proportion of queries that, when executed, yield the correct results. It is computed as the ratio of the number of accurate queries to the total number of examples in the dataset. Getting high execution accuracy is crucial for ensuring that the system's output is not only syntactically correct but also semantically meaningful and functionally accurate. This will allow this work to gauge if the approach meets the desired goals and requirements.

$$Execution\ Accuracy\ = \\ Number\ of\ SQL\ queries\ that\ return\ the\ correct\ results \\ \div Total\ number\ of\ SQL\ queries Number$$

Component matching: This metrics helps to understand which part of the SQL query is accurately predicted by the system. In this metric, each component is decomposed in predicted and ground truth query. Post that, evaluates if these two components exactly match.

This work includes the details of the data, exploratory data analysis and data preparation performed for this research purpose. WikiSQL is a dataset containing over 80,000 hand-annotated examples of SQL queries paired with their corresponding natural language descriptions and the associated SQL query corresponding to the question. This data set to include train set of 56355 records, validation set includes 8421 records and test set includes 15878 records. The length of each question is on average 25 characters and SQL query is around 30 characters.

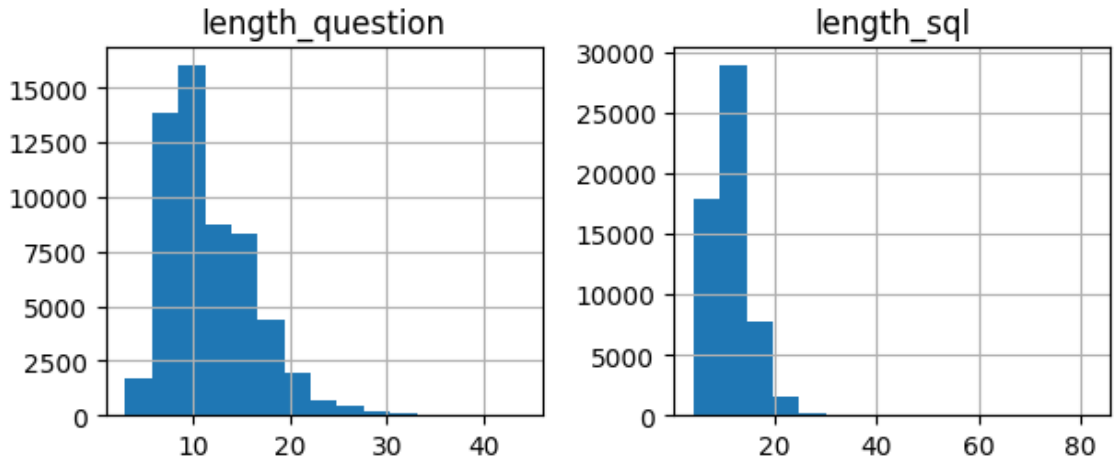

Figure 3 4.3.1 Length of question and SQL in the dataset

As per the figure 4.3: the maximum length of the SQL is 25 and question is around 30. The dataset includes Select, from and where component of the SQL grammar. There are no data related to join or other complexities of SQL grammar.

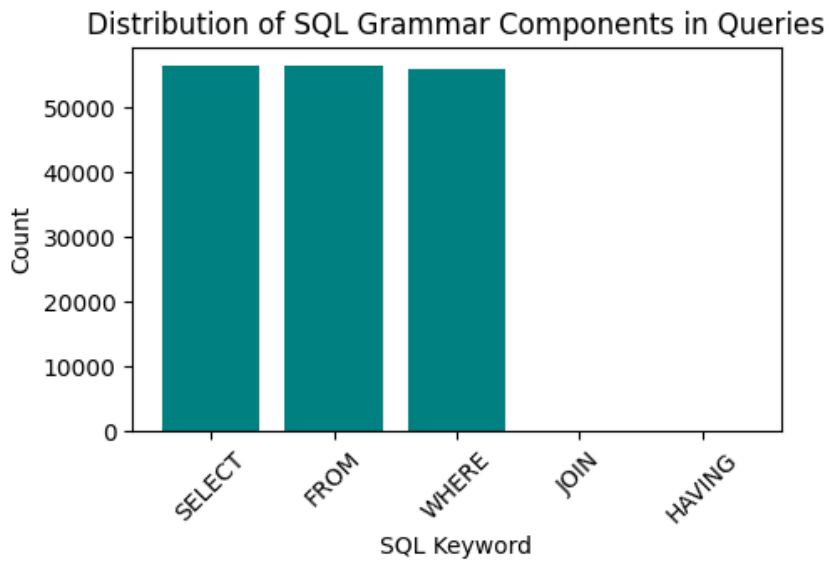

Figure 4 4.3.2 Distribution of SQL Grammer Components

As per the figure 4.3.2: the dataset includes only where clause. There are no other SQL grammar included in the dataset. The aggregators used in the dataset are count, sum, average, min and max. There are around approximately 8000 records with these aggregators.

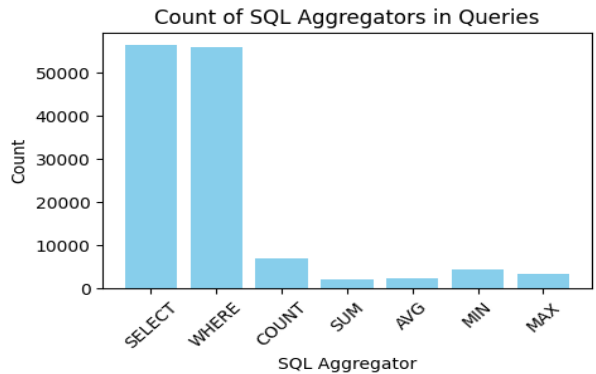

Figure 5 4.3.3 Distribution of SQL aggregator

As per the figure 4.3.3: The dataset includes Count, Sum, average, Minimum and Maximum aggregators. The data is further reviewed to understand the distribution of where clause in the dataset. Around 50% of the dataset includes single condition in the where clause. Dataset includes 10000+ records with two conditions in the where clause and less than 5000 records include three conditions in the where clause.

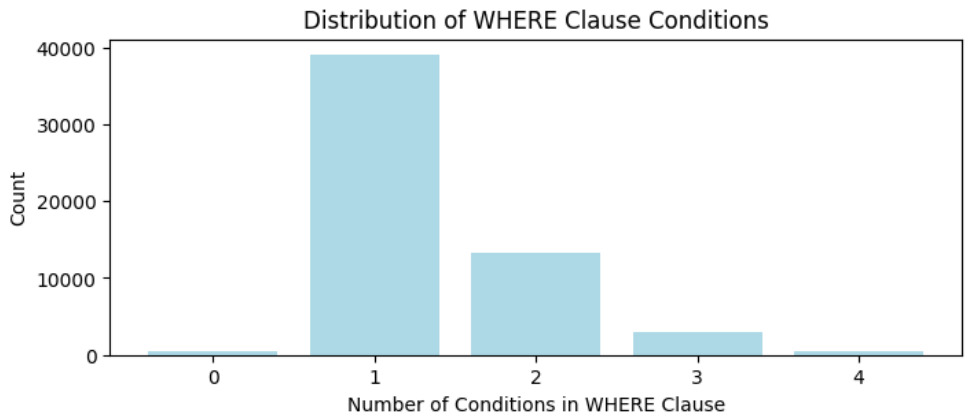

Figure 6 4.3.4 Distribution of where clause

As per the figure 4.2.4: where clause includes majorly single condition in the SQL query. Around 10000+ queries include more than one condition in the where clause and less than 5000 queries include more than 3 or 4 conditions. Thesis also reviewed the distribution of SQL operators in the where clause. 'And' operator is the majorly used operators and can be seen on more than 20000 records. 'IN' is second major operator used on 15000 records. 'OR' is the third major operator used on 10000+ records.

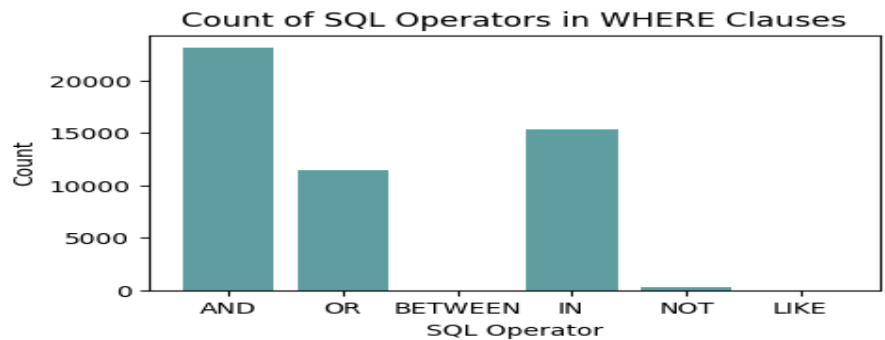

Figure 7 4.3.5 Distribution of SQL operators

As per the figure 4.3.5: the dataset includes AND, OR, IN operators for more than 10000 SQL queries. The data is further review to understand domain specific data distribution. The questions with term bigger than, larger than, more than etc are considered as domain specific data. The WikiSQL train data set includes ~7000 queries with domain specific conditions which is 14% of the train dataset. This signifies the data imbalance of normal query and query with domain specific languages. As Normalization on the dataset include user question and SQL answer, only user question is standardized by converting it to lowercase, removing special characters. SQL format is retained in the dataset so that it will be beneficial in preprocessing steps to generate the schema information. This dataset does not have any missing records. User question is tokenized into words, each word is transformed into a numerical representation. Schema of the database like table or column information is encoded into numerical format to ensure model understands the database structure. As mentioned in the 4.3 section, WikiSQL dataset includes imbalanced data for domain specific queries. To solve this problem of data imbalance in the training dataset, data balancing technique is experimented like over sampling, under sampling, synthetic data generation. Oversampling technique was used to improve the model performance in correctly predicting where clause. The main goal of this task was to increase the number of queries with domain specific questions which would provide enough example for model to learn these patterns. Identifying the domain specific questions: The first step in this experiment was to identify the domain specific questions. These queries typically included greater than or lesser than symbol in the where condition. Also, the questions included words like smaller than, bigger than, larger than, more than, less than etc. Queries with the greater than or lesser than symbols are grouped as domain specific questions. Data balancing techniques: The oversampling method included increasing the dataset with additional examples of domain specific queries. This was performed by duplicating the existing domain specific queries in the training data. Different ratio of over sampling experimented which included 1:1, 1:2 and 1:3 ratio of domain specific query to normal query within the training data. This ratio was selected based on the incrementally balancing the dataset without causing unnecessary duplication which might lead to overfitting. The frequency of normal queries is adjusted to balance the dataset. After completion of above steps, the duplicated dataset was merged with the original training dataset. This is to ensure that training data has both original and reproduced data for domain specific queries. The modified dataset is used for training the model.

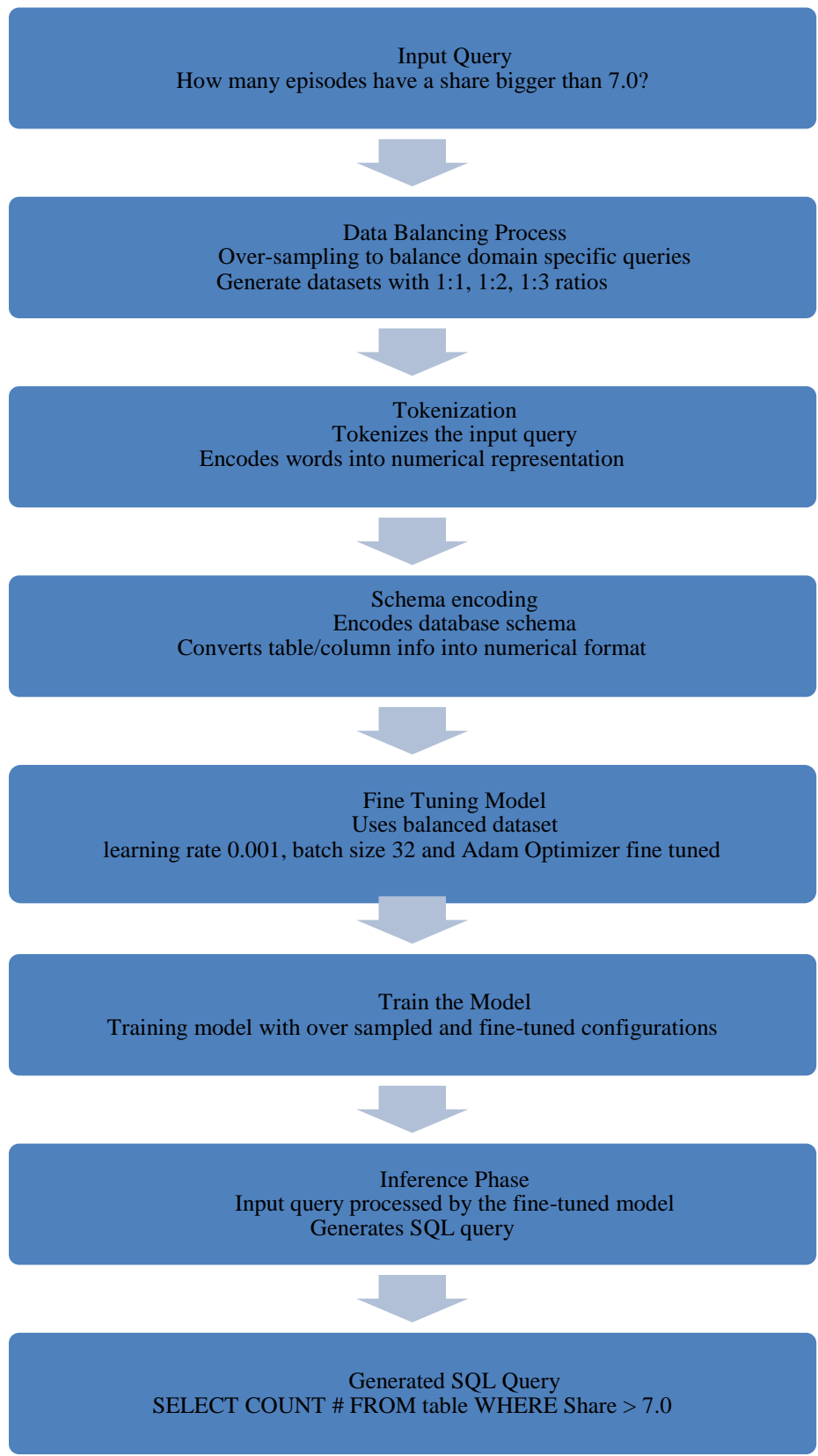

Figure 8 4.5 Combined approach of data balancing and fine tuning

Fig 4.5 shows the pictorial representation of overview of the proposed approach of fine-tuning and over-sampling for generating SQL queries from natural language inputs. The WikiSQL data set includes training, validation, and test sets. However, while building the model thesis is using the 20% of the train data set for validation of the model result. Fine tuning is another experiment performed as part of this research to improve the model's performances specifically targeting domain specific queries. Subset of data is selected focusing on queries which have domain specific languages in the where clause. Random sampling was used to ensure there is a right representation of data for the experimentation.

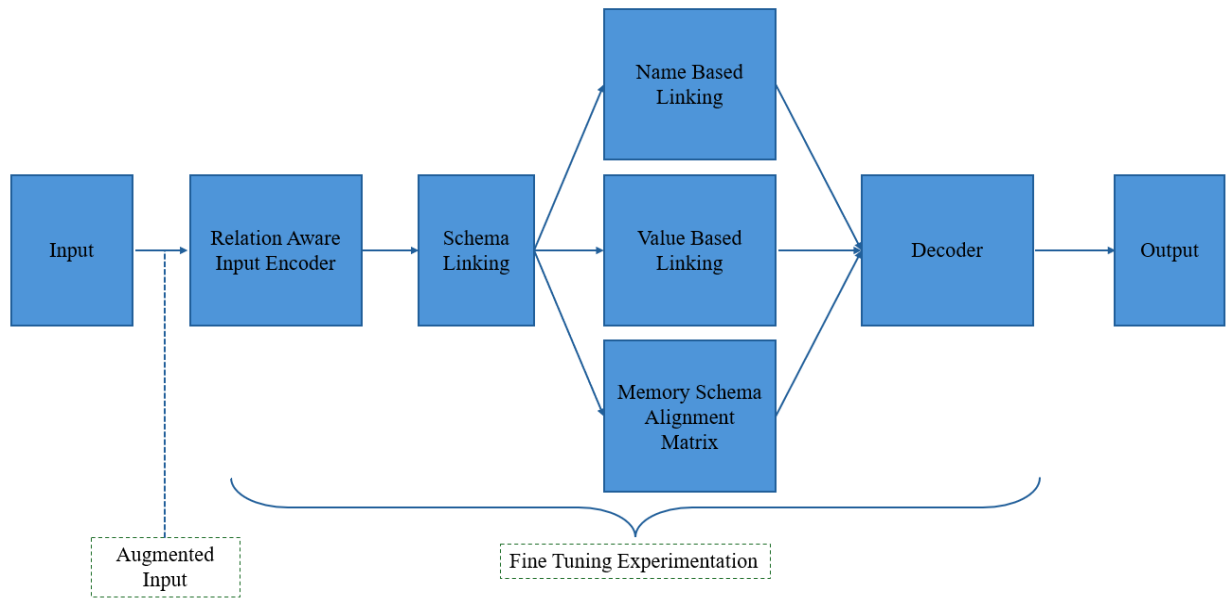

Figure 9 4.7 Proposed architecture of the model

Figure 4.7 shows the proposed architecture of the model used for this research work. Augmented input provided to the model after applying over sampling on the domain specific queries. Also, model is fine-tuned to improve the performance.

RESULTS AND DISCUSSION

The data balancing technique helped in improving the performance of the model. The performance of the model was assessed using execution accuracy, which compares the execution result of generated queries with ground truth queries. The result shown improvement in the model's accuracy when compared to the base model RAT-SQL. By providing balanced data with higher variety of domain specific query during training, the model was able to adapt to handle the different query structures. This confirmed the effectiveness of data balancing technique in text to SQL conversion system. Cross validation was used to ensure the robustness of the results. The dataset was split into multiple folds and over sampling technique was applied to each fold to measure the performance consistency. In addition to data balancing, fine tuning was experimented to enhance the model performance. This experiment was particularly aimed to refine the model's ability to understand and interpret domain specific language in the query. Subset of WikiSQL dataset was selected for fine tuning experiment focusing on domain specific queries. The model is fine-tuned on selected subset by adjusting learning rate to 0.001, batch size to 32 and Adam optimizer was used for generating the SQLs. Execution accuracy was utilized to measure the accuracy of this model. This metric calculate accuracy by reviewing if the results of executing the generated SQL query match the results of the ground truth SQL query. This metric ensures that the generated query not only syntactically matches the ground truth but also semantically retrieves the correct data.

Table 1 5.2 Comparison of performance of experimented model

| Over-Sampling Ratio with fine tuning | Execution Match Accuracy (Test) (%) | Improvement (%) |
|---|---|---|
| Baseline (RAT-SQL) [11] | 78.8 | - |
| 1:1 Ratio with fine tuning | 83.1 | +4.3 |
| 1:2 Ratio with fine tuning | 86.0 | +2.9 |
| 1:3 Ratio with fine tuning | 89.6 | +3.6 |

This section discusses the results of various oversampling ratio. 1:1 Ratio with fine tuning include applying equal number domain specific queries and normal queries. This resulted in an improvement of 4.3% from the base model. This provided the model with better opportunity to learn these patterns specific to domain and helped the model to perform better. 1:2 Ratio with fine tuning includes doubling the domain specific query. This ratio showed further improvement in the performance and increased the accuracy to 86%. This experiment showed that doubling the domain specific queries compared to normal queries, enhances the model's ability to learn these patterns even better and handle the complex where clause better. 1:3 Ratio with fine included increasing the domain specific queries three times to normal queries to provide higher level of focus learning for the model. This ensured model had plenty of examples to learn from, which significantly improved the accuracy showing increase of 3.6%. This training set helped in maintaining generalization and preventing overfitting. The use of over sampling technique to create balanced dataset ensured that model was not biased towards the common, normal queries. By providing more examples of domain specific query to the model, model was able to learn these patterns effectively and show better performance. Fine tuning with over sampling allowed the model to refine its parameters and better understand the particulars of domain specific queries. This led to improvement in the execution match accuracy, model could generalize better across different kinds of queries. The 1:3 ratio reached an ideal balance between providing enough examples of domain specific queries and maintaining a varied training set. Over sampling with fine tuning approach increased the accuracy by 10.8% from base model. In this section, the research work discusses the model performance in comparison to state-of-the-art models on WikiSQL dataset. The comparison of the logical form accuracy of this proposed model with the state-of-the-art models like HydraNet, SeaD, shows the model was able to perform better with data balancing technique. The performance of the model shows significant performance with data balancing technique against base model RAT-SQL. This shows that adding data balancing technique will benefit the model performance.

Table 2 5.3: Comparison of logical form accuracy

| Model | Logical Form Accuracy (Test) | Data balancing Technique |
|---|---|---|
| Proposed Model | 88.1% | Yes |
| SeaD (Xu et al., 2021) | 87.5% | No |
| HydraNet[4] | 86.5% | No |
| RAT-SQL [11] | 73.3% | No |

The table 5.3 shows the comparison of performance of the model logical form accuracy against state-of-the-art models. The proposed model shows better performance compared to the base model, HydraNet and SeaD. The performance of the proposed model is further analysed at each component of SQL. This is then compared against HydraNet to understand the model performance at component level. This included evaluating the performance of predicting the select column, where columns, where values, aggregators, where conditions, where operators. This helps in understanding if the model can predict components of where clause accurately for the dataset.

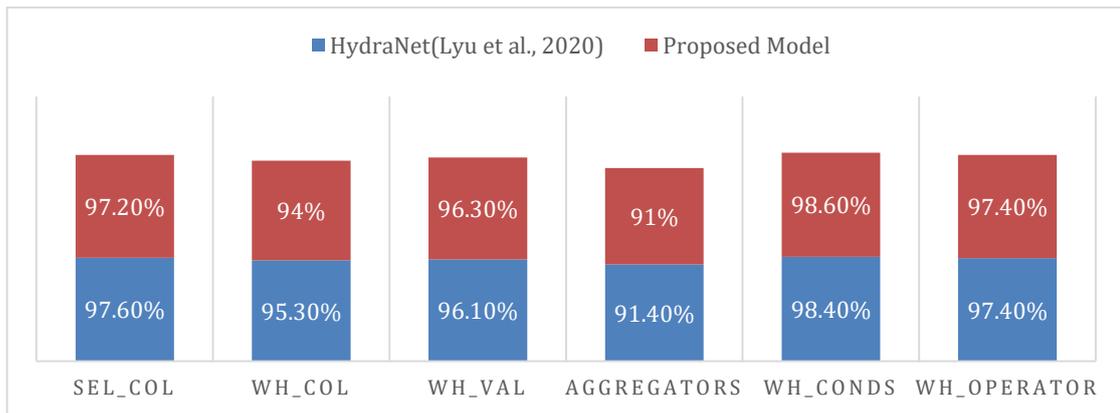

Figure 10 5.3: Comparison of each component of SQL

Figure 5.3: comparison of each component of SQL shows that proposed model's performance is closer to HydraNet specially in predicting select columns, where values, where conditions and operators. Though the overall logical accuracy is at 88.1%, model can perform better at predicting individual components of the SQL query. This difference can be attributed to the difficulty in achieving the logical consistency across all components of query. High component level accuracy indicates that model can perform better at individual tasks but there needs to be improvement on integrating these components to a logical SQL query. Though model showed better accuracy at individual component of SQL, the logical accuracy indicates the correctness of the SQL query is low. The discrepancy indicates that model can accurately identify and process the individual query component but fails at integrating them into a logical SQL query. Reason for dip in accuracy is reviewed when compared to component level accuracy. There are couple of major reasons for this behaviour; 18% of the queries predicted inaccurately due to component interdependency. This is happening when model selected the correct column but applied incorrect aggregation. E.g.: instead of applying count as aggregator model applying sum as aggregator on a selected column. 12% of the subtasks are not integrated accurately. Error occurred when combining the accurately processed components into logical query. E.g.: switching sel_col with wh_col while integrating the query. This can be improved by implementing iteratively analysing and correcting the errors to refine the model performance.

Data Availability:

The data that support the findings of this study are openly available in WikiSQL

CONCULSIONS AND RECOMMENDATIONS

In summary, the implementation of oversampling and fine-tuning techniques on this dataset enhances the accuracy of the system when dealing with the domain specific user questions. These improvements demonstrate the value of proposed data handling and model training approaches enhances the model performance in conversion of text to SQL. The enhanced technique of dataset oversampling, and model fine-tuning have effectively mitigated the challenges caused by data imbalance in the training set. These modifications have directly translated into enhanced model performance, demonstrating their efficiency in the context of natural language to SQL translation tasks. This study shows the importance of proposed data handling and training approaches in the development of machine learning models for specific tasks. Using execution match metrics to calculate accuracy provided a robust evaluation of the model's performance on domain-specific queries. Through careful analysis and experimentation, it was determined that a 1:3 over-sampling ratio yielded the best results, with an execution match accuracy of 89.6%. This optimal ratio ensured a balanced dataset, allowing the model to learn effectively from enough domain-specific examples. Future work will explore combining over-sampling with other data balancing techniques, such as synthetic data generation and data augmentation, to further enhance model performance. Additionally, exploring the integration of advanced optimization algorithms and alternative model architectures could provide further improvements in handling domain-specific SQL translation tasks. Implementation of oversampling and fine tuning was technically challenging due to the size of the data and resource constraints. Hence, the small set of whole data was used to experiment these enhancements locally. Expanding this proposed model to entire dataset can also be perused as future work.